\documentclass[aps,reprint,preprintnumbers,amsmath,amssymb,superscriptaddress,prl]{revtex4-1}

\usepackage{graphicx,color}

\usepackage{dcolumn}

\usepackage{bm}

\usepackage[dvipdfmx]{hyperref}

\begin{document}


\title{Quasi-one-dimensional Bose-Einstein Condensation\\in the Spin-1/2 Ferromagnetic-leg Ladder 3-I-V}

\author{Y. Kono}
\email{k-yohei@issp.u-tokyo.ac.jp}
\affiliation{Institute for Solid State Physics, the University of Tokyo, Kashiwa 277-8581, Japan}
\author{S. Kittaka}
\affiliation{Institute for Solid State Physics, the University of Tokyo, Kashiwa 277-8581, Japan}
\author{H. Yamaguchi}
\affiliation{Department of Physical Science, Osaka Prefecture University, Osaka 599-8531, Japan}
\author{Y. Hosokoshi}
\affiliation{Department of Physical Science, Osaka Prefecture University, Osaka 599-8531, Japan}
\author{T. Sakakibara}
\affiliation{Institute for Solid State Physics, the University of Tokyo, Kashiwa 277-8581, Japan}

\date{\today}

\begin{abstract}
Quantum criticality of the spin-1/2 ferromagnetic-leg ladder 3-I-V [=3-(3-iodophenyl)-1,5-diphenylverdazyl] has been examined with respect to the antiferromagnetic to paramagnetic phase transition near the saturation field $H_{c}$. The phase boundary $T_{c}(H)$ follows the power-law $T_{c}(H)\propto H_{c}-H$ for a wide temperature range. This characteristic behavior is discussed as a quasi-one-dimensional (quasi-1D) Bose-Einstein condensation, which is predicted theoretically for weakly coupled quasi-1D ferromagnets. Thus, 3-I-V provides the first promising candidate for this attractive prediction.  
\end{abstract}



\maketitle

Low dimensionality---one or two dimensions (1D or 2D)---plays an essential role in various exotic quantum phenomena such as a Tomonaga--Luttinger liquid (TLL), which is related to 1D spinon excitations~\cite{Giamarchi}, and a Haldane state~\cite{PhysRevLett.50.1153} and Kosterlitz-Thouless transition~\cite{0022-3719-6-7-010}, which are related to topological phases. However, in real quantum magnets, the effects of three-dimensional (3D) interactions are inevitable so that they are always \lq\lq{}quasi\rq\rq{}low dimensional. 

Since the universality of a quantum phase transition reflects the space dimensionality of the system~\cite{2011qpbookS,0034-4885-66-12-R01}, the critical behavior near the quantum critical point (QCP) in a quasi-low-dimensional magnet should belong to the 3D universality class. A notion of Bose-Einstein condensation (BEC) has often been applied to describe a certain universality class of magnetic-field-induced QCP; 3D $XY$ antiferromagnetic (AFM) ordering can be mapped onto the condensate of lattice-gas bosons~\cite{RevModPhys.86.563,giamarchi2008bose}. Realization of the 3D BEC QCP has been intensively studied in spin dimer systems such as TlCuCl$_{3}$~\cite{JPSJ.77.013701} and BaCuSi$_{2}$O$_{6}$~\cite{PhysRevB.72.100404,*sebastian2006dimensional}. As quasi-low-dimensional magnets, the spin-1/2 triangular-lattice antiferromagnet Cs$_{2}$CuCl$_{4}$~\cite{PhysRevLett.95.127202,*PhysRevLett.96.189704}, the spin-1/2 AFM two-leg ladders (Cu$_{7}$H$_{10}$N)$_{2}$CuBr$_{2}$~\cite{PhysRevLett.111.106404} and (Cu$_{5}$H$_{12}$N)$_{2}$CuBr$_{4}$~\cite{PhysRevB.79.020408}, and the spin-1/2 alternating AFM spin chain Cu(NO$_{3}$)$_{2}\cdot$2.5H$_{2}$O (copper nitrate)~\cite{PhysRevB.91.060407} are representative.
 
Moving away from the QCP, the weaker 3D interactions are effectively masked, and low-dimensional characteristics can arise. One of the attractive issues of this type of dimensional crossover is the theoretical proposal for a BEC in quasi-low-dimensional magnets including ferromagnetic (FM) chains (quasi-1D) or planes (quasi-2D); the power law of the critical temperature near the saturation field $H_{c}$, $T_{c}(H)\sim|H_{c}-H|^{1/\phi}$, can exhibit crossovers from $\phi\,=\,3/2$ with the 3D BEC universality class to $\phi\simeq1$ with quasi-1D or quasi-2D ones as it moves away from the QCP~\cite{PhysRevB.75.134421}. Although the 3D BEC exponent has widely been investigated as mentioned above, there exist few experimental tests for such dimensional crossovers. The quasi-2D BEC case has been reported for the spin-1/2 2D $XXZ$ ferromagnet K$_{2}$CuF$_{4}$~\cite{PhysRevB.95.174406}, which is referred to as one of the candidates in Ref.~\cite{PhysRevB.75.134421}. On the other hand, candidates for the quasi-1D BEC case remain to be explored. 

\begin{figure}[b]
\begin{center}
\includegraphics[width=0.75\linewidth]{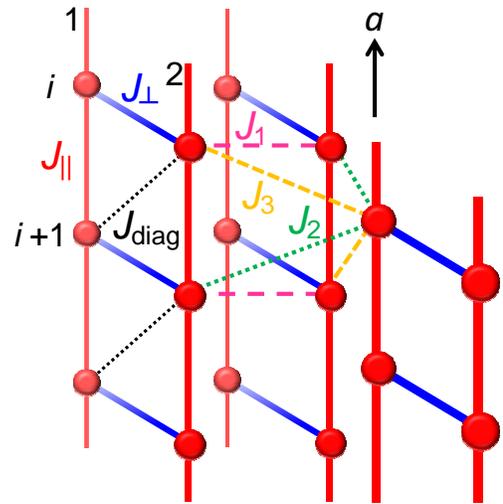}
\caption{Schematic of a part of the dominant intermolecular interactions of 3-I-V, as represented in Ref.~\cite{PhysRevB.91.125104}. Each sphere corresponds to an $S\,=\,\frac{1}{2}$ spin (see the text).}\label{f0}
\end{center}
\end{figure}

Recently, a new type of quasi-1D quantum magnet has been synthesized using verdazyl radical-based molecules, each of which carries an $S\,=\,\frac{1}{2}$ spin|spin-1/2 FM-leg ladder~\cite{doi:10.7566/JPSJ.83.033707,PhysRevLett.110.157205,PhysRevB.89.220402,PhysRevB.91.125104}.
The Hamiltonian of a typical two-leg spin ladder in a magnetic field can be described as  
\begin{eqnarray}
\mathcal{H}&=&J_{||}\sum_{i,\alpha}\bm{S_{i,\alpha}\cdot S_{i+1,\alpha}}+J_{\perp}\sum_{i}\bm{S_{i,1}\cdot S_{i,2}}\nonumber\\&&-g\mu_{B}H\sum_{i,\alpha}S^{z}_{i,\alpha},\label{ladder}
\end{eqnarray}
where $J_{||}$ is the interaction along each leg ($\alpha=1,2$), $J_{\perp}$ is the rung interaction between the legs, $g$ is the $g$ factor, and $\mu_{B}$ is the Bohr magneton (see Fig.~\ref{f0}). The FM-leg case corresponds to $J_{||}\,<\,0$ and $J_{\perp}\,>\,0$. Previously, we reported that the 3D BEC exponent was observed on one of the FM-leg ladders 3-Br-4-F-V [=3-(3-bromo-4-fluorophenyl)-1,5-diphenylverdazyl] near the lower critical and saturation field~\cite{PhysRevB.96.104439,note2}. 3-Br-4-F-V is a strong-rung ladder ($|J_{||}/J_{\perp}|\,<\,1$), whereas the other two 3-Cl-4-F-V [3-(3-chloro-4-fluorophenyl)-1,5-diphenylverdazyl]~\cite{PhysRevLett.110.157205} and 3-I-V [=3-(3-iodophenyl)-1,5-diphenylverdazyl]~\cite{PhysRevB.91.125104} are strong-leg ladders ($|J_{||}/J_{\perp}|\,>\,1$). Thus, the effect of the difference in the intraladder interactions for the critical phenomena near the critical field is of great interest.

In this study, we show that the spin-1/2 FM-leg ladder 3-I-V [=3-(3-iodophenyl)-1,5-diphenylverdazyl] is a promising candidate for quasi-1D BEC. The phase boundary of the 3D ordering, $T_{c}(H)$, near the saturation field $H_{c}$ was precisely determined by several experimental methods. The obtained $T_{c}(H)$ is in accordance with the power law $T_{c}(H)\propto(H_{c}-H)$ over a wide temperature range. The possibility of quasi-1D BEC is discussed based on the dominant spin interactions of 3-I-V. 

\begin{figure}[b]
\begin{center}
\includegraphics[width=\linewidth]{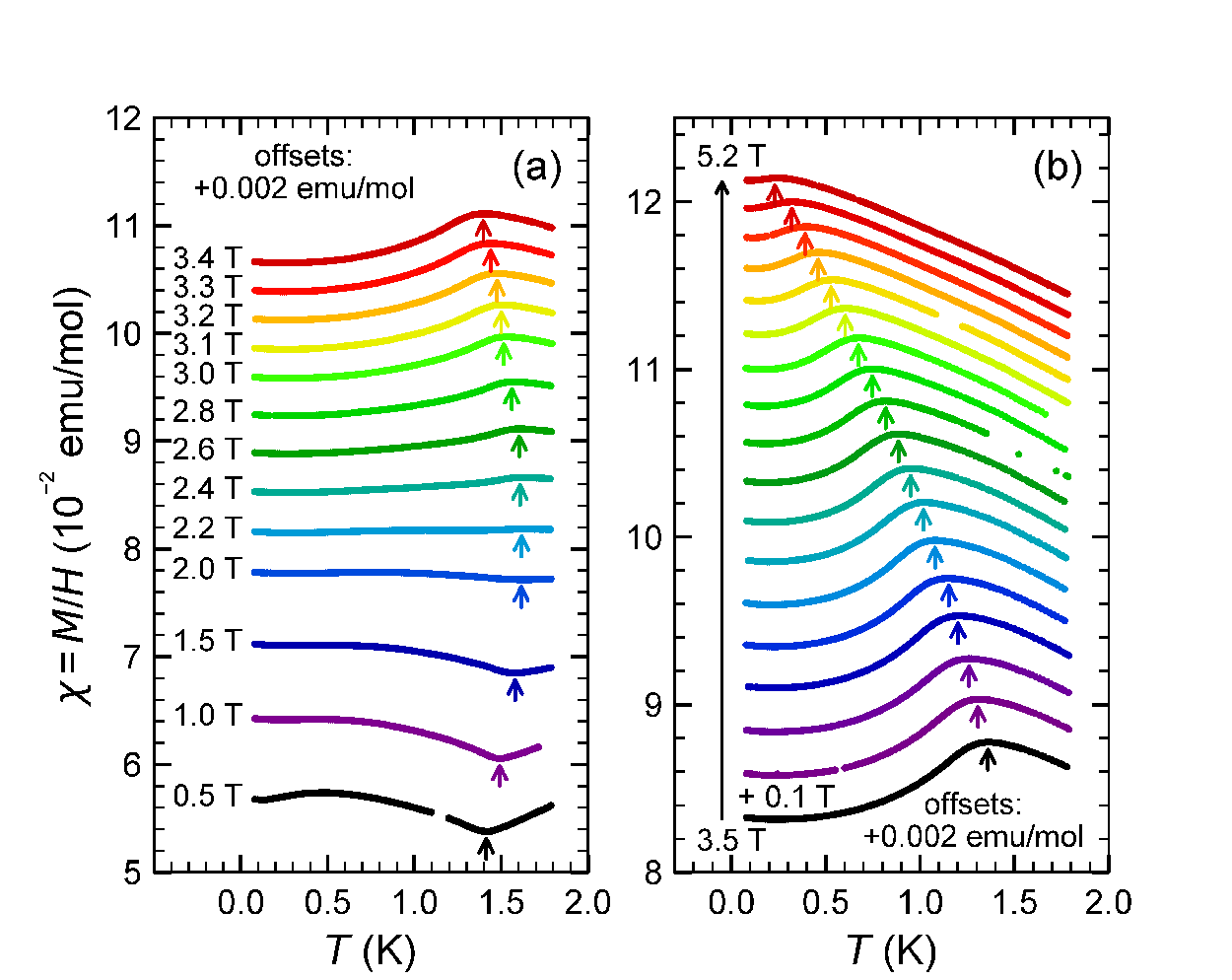}
\caption{Temperature dependence of the magnetic susceptibility $\chi\,=\,M/H$ in several magnetic fields (a) from 0.5\,T to 3.4\,T and (b) from 3.5\,T to 5.2\,T with 0.1\,T steps. Each curve is shifted by $+0.002$\,emu/mol for clarity. The arrows indicate the 3D ordering temperature $T_{c}$ at which $\chi$ has a cusp-like minimum or maximum.}\label{f1}
\end{center}
\end{figure}

Predominant intermolecular interactions of 3-I-V have been predicted as a FM-leg spin ladder along the $\bm{a}$ axis by \emph{ab initio} molecular orbital (MO) calculations~\cite{doi:10.7566/JPSJ.83.033707,PhysRevB.91.125104}, and the intraladder interactions were estimated at $J_{||}/k_{B}\,=\,-11.6\,$K and $J_{\perp}/k_{B}\,=\,5.8\,$K by comparison of the experiments with quantum Monte Carlo simulations~\cite{doi:10.7566/JPSJ.83.033707}. 
Although the isotropic FM-leg spin ladder has a spin gap as long as $J_{\perp}\,\neq\,0$~\cite{PhysRevB.67.064419,*PhysRevB.70.014425}, 3-I-V exhibits 3D ordering at zero field, and the 3D ordering phase reaches the saturation field near 5.5\,T~\cite{PhysRevB.91.125104}. The 3D ordering has been attributed to the frustrated intra and interladder couplings predicted by the MO calculations; the diagonal interactions $J_{\mathrm{diag}}\,<\,0$ and the interactions $J_{1}\,>\,0$, $J_{2}\,<\,0$, and $J_{3}\,<\,0$, which form triangles, as illustrated in Fig.~\ref{f0}. These interactions were estimated at approximately $0.1J_{\perp}$~\cite{PhysRevB.91.125104}. Note that MO calculations have been proven effective to evaluate predominant intermolecular interactions of verdazyl radical compounds~\cite{doi:10.7566/JPSJ.83.033707, doi:10.7566/JPSJ.82.043713, PhysRevB.88.174410, PhysRevB.88.184431}. 

\begin{figure}[b]
\begin{center}
\includegraphics[width=\linewidth]{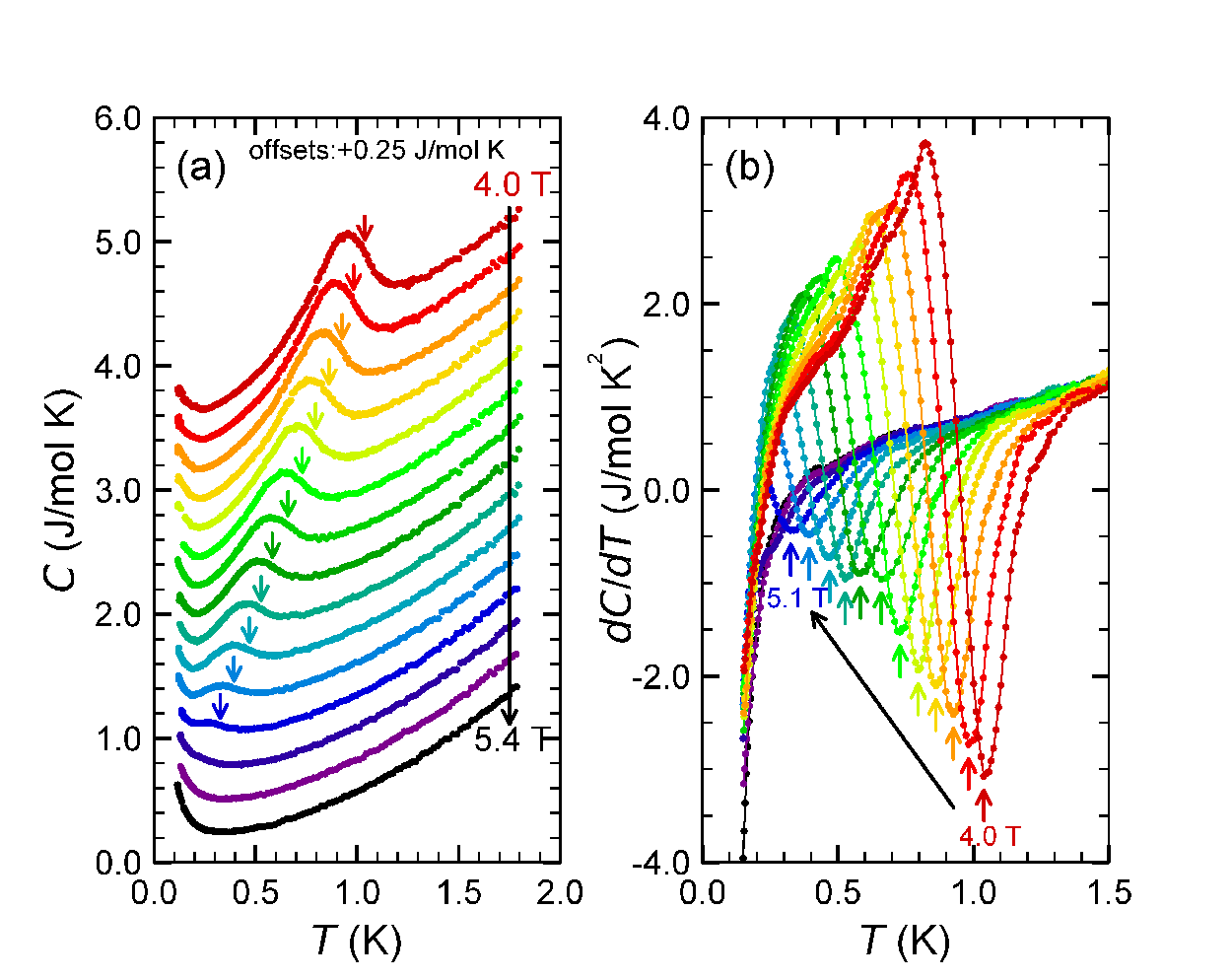}
\caption{(a) Temperature dependence of the heat-capacity $C$ and (b) its temperature derivative $dC/dT$ in several magnetic fields from 4.0\,T to 5.4\,T with 0.1-T steps. Each curve in (a) is shifted by $+0.25$\,J\,mol$^{-1}$\,K$^{-1}$ for clarity. The arrows indicate the 3D ordering temperature $T_{c}$ at which $C$ shows an inflection just beyond the peak.}\label{f2}
\end{center}
\end{figure}

\begin{figure}[t]
\begin{center}
\includegraphics[width=0.95\linewidth]{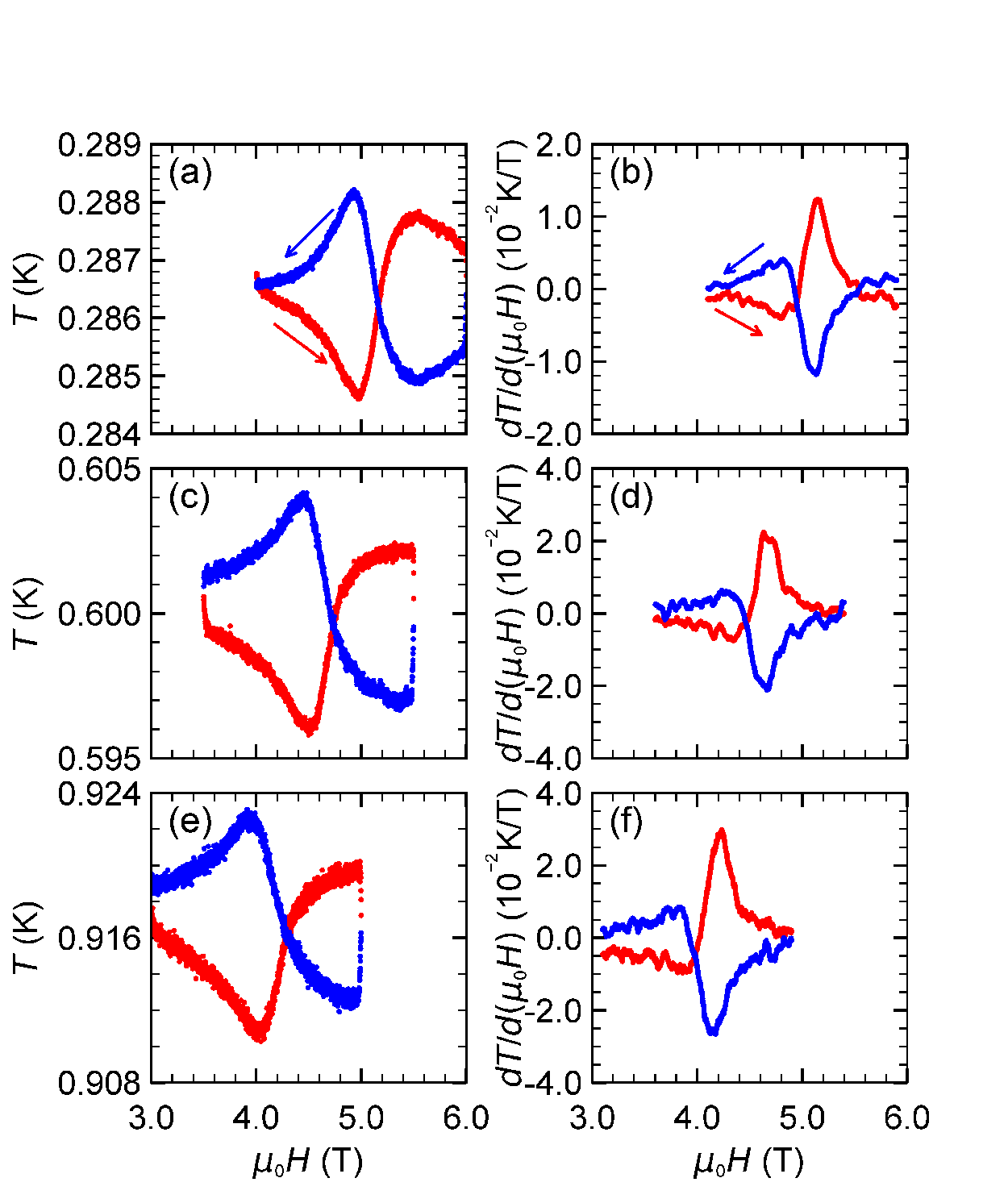}
\caption{(a), (c), and (e) show magnetocaloric-effect curves at several fixed bath temperatures. (b), (d), and (f) show $dT/d(\mu_{0}H)$ obtained from (a), (c), and (e), respectively. THe red or lighter gray (blue or darker gray) curves denote the up- (down-)sweep measurements.}\label{f3}
\end{center}
\end{figure}

\begin{figure}[t]
\begin{center}
\includegraphics[width=0.95\linewidth]{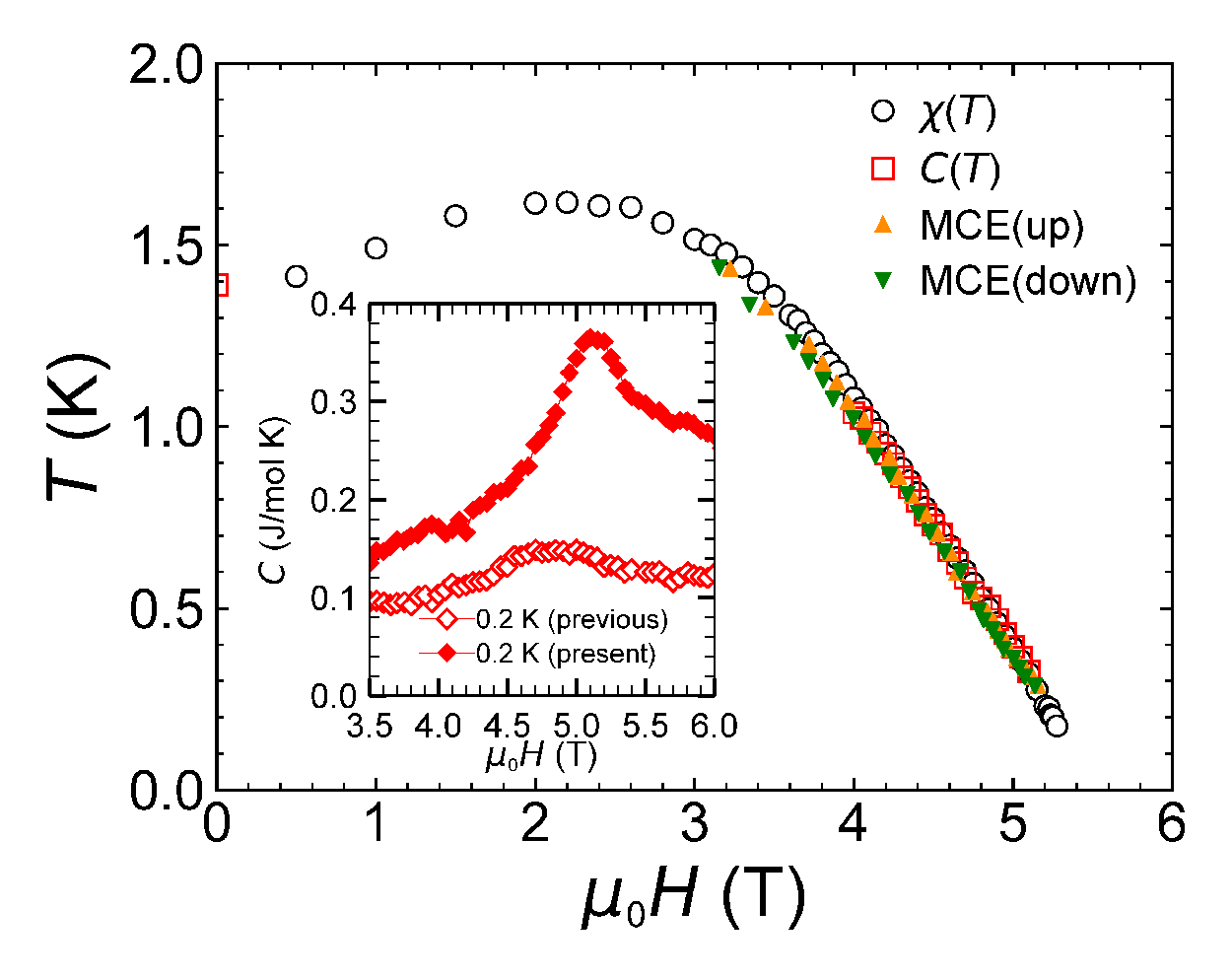}
\caption{Phase boundary of the 3D ordering, $T_{c}(H)$, determined from the present measurements. The open circles and squares show $T_{c}$ obtained from $\chi(T)$ and $C(T)$, respectively. Up and down triangles show $T_{c}$ obtained from $dT/d(\mu_{0}H)$ of the up- and down-sweep MCE data, respectively. The inset: magnetic-field dependence of $C$ near the saturation field in the present (the closed diamonds) and previous (the open diamonds, Ref.~\cite{PhysRevB.91.125104}) measurements.}\label{f4}
\end{center}
\end{figure}

To extract the critical exponent $\phi$ of 3-I-V near the saturation field $H_{c}$, we performed a precise determination of the 3D ordering phase boundary using three methods: magnetization, specific-heat, and magnetocaloric-effect (MCE) measurements.
Single-crystal samples of 3-I-V were synthesized as described in Ref.~\cite{doi:10.7566/JPSJ.83.033707}. dc magnetization measurements were performed by a force magnetometer~\cite{JJAP.33.5067} on 2.52-mg randomly oriented samples. Specific-heat measurements were carried out by the standard quasiadiabatic heat-pulse method on the same samples. The MCE was measured by up and down magnetic-field sweepings at 50-80 mT/min with fixed bath temperatures using a 0.42-mg crystal from the same samples. For all measurements, magnetic fields up to 6\,T were applied perpendicular to the $\bm{a}$ axis (perpendicular to the leg direction).  

Figure \ref{f1} shows the temperature dependence of the magnetic susceptibility $\chi\,=\,M/H$ in several magnetic fields from 0.5 to 5.2\,T. There exists a cusplike minimum or maximum in each curve as reported in Ref.~\cite{PhysRevB.91.125104}. This cusplike anomaly is typically observed in model materials discussed in the context of BEC~\cite{JPSJ.77.013701,PhysRevB.69.020405,PhysRevB.95.174406} and is predicted by theoretical calculations~\cite{PhysRevLett.84.5868,PhysRevB.73.014433}. In the case of two-leg spin ladder systems, extrema associated with a crossover to the TLL regime have often been observed above 3D ordering temperatures~\cite{PhysRevLett.108.097201,PhysRevB.91.060407,PhysRevB.83.054407,PhysRevLett.84.5399,PhysRevLett.87.206407}, but 3-I-V has no such anomalies, which is attributed to the interladder interactions. Thus, the cusplike anomaly indicates the position of the 3D ordering temperature $T_{c}$. 

In Fig.~\ref{f2}(a), the temperature dependence of the specific-heat $C$ in several magnetic fields from 4.0 to 5.1\,T exhibits a peak anomaly. The upturn behavior at low temperatures is attributed to nuclear Schottky contributions from $^{1}$H, $^{127}$I, and $^{14}$N. Although the peak anomaly itself was defined as $T_{c}$ in a previous report~\cite{PhysRevB.91.125104}, an inflection point beyond the peak anomaly of $C$ is a more plausible way to define $T_{c}$ because it shows excellent agreement with $T_{c}$ as determined from $\chi(T)$ and the MCE measurements. Therefore, we assign the sharp trough in $dC/dT$ to $T_{c}$ as shown in Fig.~\ref{f2}(b).

Figures~\ref{f3}(a), \ref{f3}(c), and \ref{f3}(e) show the MCE curves at fixed bath temperatures $\sim$286.5, $\sim$60, and $\sim$918\,mK, respectively. The up-sweep and down-sweep curves on each panel appear to be almost vertically symmetric with each other. Such behavior indicates that the MCE measurements were performed under equilibrium conditions~\cite{PhysRevLett.96.077204,RevModPhys.86.563}. Under these conditions, the first derivative of $T(H)$ exhibits a sharp peak (trough) in the up- (down-) sweep curves as shown in Figs.~\ref{f3}(b), \ref{f3}(d), and \ref{f3}(f). The phase boundary $T_{c}(H)$ can be determined from the positions of the peak (trough) anomalies in a similar manner to that used in the MCE measurements of other systems under similar conditions~\cite{PhysRevLett.96.077204,PhysRevB.77.214441,PhysRevB.79.100409}. Note that the temperature difference between the sample and the bath temperature $\Delta T(H)$ at the equilibrium conditions behaves like
\begin{equation}
\Delta T(H) = -\frac{T}{\kappa}\dot{H}\left(\frac{\partial M}{\partial T}\right)_{H}, \label{eq:equi}
\end{equation}
where $\kappa$ is the thermal conductivity between the sample and the bath and $\dot{H}$ is the sweep rate~\cite{RevModPhys.86.563}. This implies that the extrema in $\chi(T)[\,=\,M(T)/H]$ should be associated with the inflection points of the MCE curves. This fact supports the agreement of the definitions of $T_{c}$ in the $\chi(T)$ and MCE results.

The 3D ordering temperatures $T_{c}(H)$, determined from the present measurements are summarized in Fig.~\ref{f4}. All the definitions of $T_{c}$ are in excellent agreement with each other, so that they give the exact phase boundary of 3-I-V. We can also refer to the additional phase discussed in the previous report~\cite{PhysRevB.91.125104}. In the inset of Fig.~\ref{f4}, the magnetic-field dependence of the specific heat for the present sample is compared with previous measurements. The present measurements show only a single sharp peak, different from the broad peak shown in the previous measurements. This implies that the broad peak in the previous data may arise from a collapse of the sharp peak rather than an overlap of two phase transitions. Such a collapse of a peak in specific-heat measurements indicates an effect of disorder~\cite{PhysRevB.84.174414} so that the sharper peak in the present results is attributed to the improvement of the sample quality. Moreover, there exist no additional anomalies in $\chi(T)$ and $C(T)$ near the additional phase boundary defined previously. Thus, in fact, we conclude that the additional phase does not exist.

\begin{figure}[t]
\begin{center}
\includegraphics[width=\linewidth]{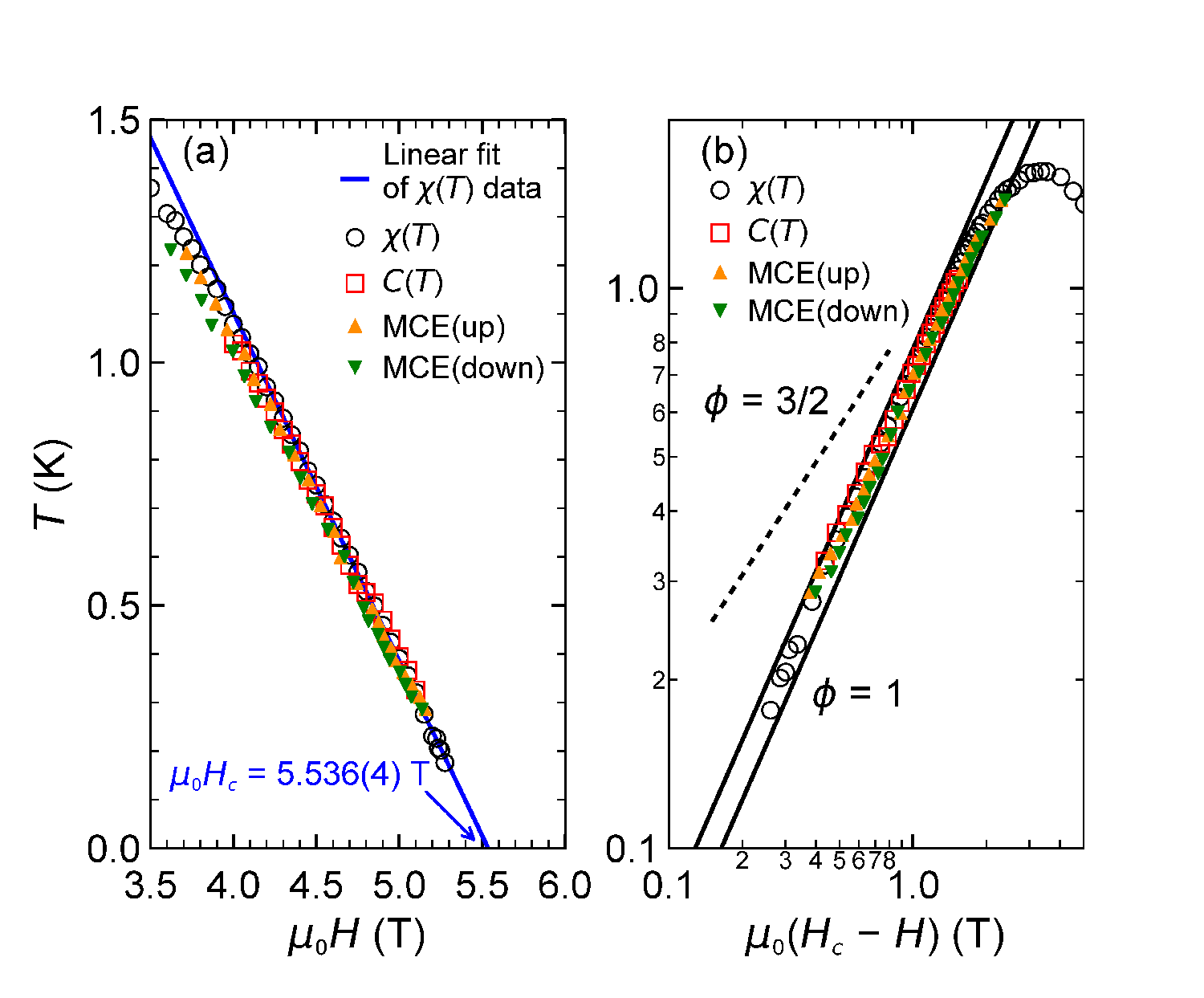}
\caption{(a) Enlarged plot of the main panel of Fig.~\ref{f4} near the saturation field $H_{c}$. Solid line is the linear-fitting line of $T_{c}(H)$ from the $\chi(T)$ data below 1\,K. Extrapolating the linear-fitting line to zero temperature yields $H_{c}$\,=\,5.536(4)\,T. (b) Log-log plot of $T_{c}(H)$ vs $\mu_{0}(H_{c}-H)$. The solid lines correspond to $T_{c}(H)\propto H_{c}-H$. The dotted line corresponds to $T_{c}(H)\propto (H_{c}-H)^{2/3}$ for comparison.}\label{f5}
\end{center}
\end{figure}

The quantum criticality of the phase boundary near the saturation field is represented in Fig.~\ref{f5}. As indicated in Fig.~\ref{f5}(a), $T_{c}(H)$ from the $\chi(T)$ data below 1\,K is well reproduced by the linear fitting of the data, yielding the critical field $\mu_{0}H_{c}\,=\,5.536(4)\,$T. Figure~\ref{f5}(b) is a log-log plot of $T_{c}(H)$ vs $\mu_{0}(H_{c}-H)$ for all the data of $T_{c}(H)$. It demonstrates that all the definitions of $T_{c}$ below 1\,K are consistent with the line corresponding to the critical exponent $\phi\,=\,1$ (the solid lines), clearly distinguished from $\phi\,=\,\frac{3}{2}$ of the 3D BEC case (the dotted line). The quasi-1D or quasi-2D BEC predicted in Ref.~\cite{PhysRevB.75.134421} is hence expected to be realized in 3-I-V.

The quasi-1D case is compatible with the spin interactions of 3-I-V predicted by the MO calculations. As discussed in Ref.~\cite{PhysRevB.75.134421}, the quasi-1D BEC exponent $\phi\,=\,1$ can be found if effective interactions between magnons are sufficiently small near QCP. In 3-I-V, quasiparticle (magnon) excitation near the QCP ($H_{c}$) could be principally derived from the transition from the triplet to singlet state on each rung because the rung interactions $J_{\perp}$ are antiferromagnetic. The magnons on each two-leg ladder interact through the frustrated $J_{1}$-$J_{2}$-$J_{3}$ interactions as described above. Such frustration with the opposite signs of interactions could suppress the effective interactions between those magnons, so that the predominant FM-leg interactions are relatively enhanced to cause the quasi-1D BEC exponent $\phi\,=\,1$. One can suspect that the ladder-type interactions cause the 2D characteristics, but it is required for the quasi-2D BEC that a 2D plane consists of ferromagnetic interactions as described in Ref.~\cite{PhysRevB.75.134421}. Thus, the quasi-2D BEC case is not suitable for the present conditions. Crossover to the 3D BEC exponent is not observed in the temperature range of this study although the interladder interactions are on the order of $0.1J_{\perp}$. This is also attributed to the effect of the frustration such that the crossover would be found below 0.1\,K. Since these arguments are only on the basis of the MO calculations, inelastic neutron-scattering measurements would be needed to estimate more accurate exchange parameters. 

To summarize, we have examined the quantum critical phenomena near the saturation field ($H_{c}$) on the spin-1/2 FM-leg ladder 3-I-V. The phase boundary of the 3D ordering state near $H_{c}$ was precisely determined by magnetization, specific-heat, and MCE measurements. All definitions of the 3D ordering temperatures $T_{c}$ are in excellent agreement with each other. The obtained phase boundary shows the linearity of the power-law $T_{c}(H)\propto H_{c}-H$ below 1\,K. The characteristic behavior would be caused by quasi-1D BEC with the predominant ferromagnetic interactions proposed by Ref.~\cite{PhysRevB.75.134421}, which could be enhanced by the interladder frustration. Thus, spin-1/2 FM-leg ladders are promising for investigating the relationship between low dimensionality and BEC physics in quantum magnets.   

This work was supported in part by KAKENHI Grants No. 16J01784, No. 15K05158, No. 17H04850, No. 15H03695, and No. 15H03682 from JSPS. The sample preparation of 3-I-V was performed at Osaka Prefecture University. The magnetization, specific-heat, and MCE measurements were conducted at the Institute for Solid State Physics, the University of Tokyo.

\begin{thebibliography}{40}
\bibitem [{not()}]{note2}%
  \BibitemOpen
  \href@noop {} {}\bibinfo {note} {Note that the critical exponent in
  Ref.~\cite{PhysRevB.96.104439} is defined as $\nu\,=\,1/\phi$}\BibitemShut
  {NoStop}%
\end{thebibliography}
%

\end{document}